
\def\e{{\rm e}}
\def\p{\medskip}
\def\fr{{\partial f\over \partial r}}
\def\fth{{\partial f\over \partial \theta}}
\def\Wr{{\partial W\over \partial r}}
\def\Wth{{\partial W\over \partial \theta}}
\def\Sr{{\partial S\over \partial r}}
\def\Sth{{\partial S\over \partial \theta}}
\def\Yr{{\partial Y\over \partial r}}
\def\Yth{{\partial Y\over \partial\theta}}
\def\hr{{\partial h\over \partial r}}
\def\hth{{\partial h\over \partial\theta}}

\def\fx{{\partial f\over \partial x}}
\def\hx{{\partial h\over \partial x}}

\def\turokz{1}
\def\dine{2}
\def\bdt{3}
\def\arnold{4}
\def\klinkm{5}
\def\mog{6}
\def\ambj{7}
\def\damg{8}
\def\pd{9}
\def\markt{10}
\def\menpb{11}
\def\nambu{12}
\def\mogtan{13}
\def\vort{14}
\def\meprd{15}
\magnification=\magstep1
\overfullrule=0pt
\font\large=cmr10 scaled \magstep 2
\centerline{\large Baryon Number Violating Transitions in String Backgrounds}
\bigskip
\bigskip
\centerline{Warren B. Perkins}
\bigskip
\centerline{Department of Physics}
\centerline{University of Wales, Swansea}
\centerline{Singleton Park}
\centerline{Swansea, SA2 8PP}
\vskip 2.0in
\centerline{Abstract}

We construct field configurations that interpolate between string background
states of differing baryon number. Using these configurations we estimate the
effect of the background fields on the energy barrier separating different
vacua.  In the background of a superconducting GUT string the energy barrier is
increased, while in
an electroweak string background or the electroweak layer of a
non-superconducting string the energy
barrier is reduced. The energy barrier depends sensitively on both the
background gauge and scalar fields.
\vfil\eject
\noindent{\bf 1) Intoduction}
\p
The realisation that electroweak baryon number violation can occur
at an appreciable rate in the early universe has raised the hope of
explaining the observed baryon asymmetry of the universe with TeV
scale physics. Most models rely on a first order phase transition
that proceeds via the nucleation and subsequent growth of true
vacuum bubbles[\turokz]. The order of this phase transition is not certain and
it is not
clear that it is sufficiently strongly first order for these mechanisms to
work[\dine]. This
provides an incentive to produce scenarios that do not rely on the
order of the electroweak transition.
Just such a method has been suggested by Davis, Brandenberger and
Trodden[\bdt] who make use of the decay of a preexisting string network to
provide
the departure from thermal equilibrium. Just as in most methods based on
bubble nucleation, an extension of the standard model is required to increase
CP violation, this is provided by extending the Higgs sector of the theory. The
coupling of the extra Higgs field is such as to generate a chemical potential
in regions that are entering or leaving the core of the string. This chemical
potential
leads to the generation of antibaryons at the leading edge of the string and
baryons on its
trailing edge. The decay of the string network causes a greater volume of space
to leave the
string core region than enter it, leading to a net production of baryons.
As the antibaryons are produced on the leading edge of the
string, they enter the core region shortly after production and baryon
production is enhanced if
they can decay before the string leaves them behind.
If the baryon asymmetry is to survive, the rate of baryon number violation
outside the
string should be small, that is we are in an epoch when the sphaleron energy is
large
compared to the temperature so that the sphaleron transition rate [\arnold]
$$
\Gamma_{\Delta B} =\gamma (\alpha_{wk} T)^{-3} M_W^7 \e^{(-4\pi
\nu(T)/g_{wk}T)}
$$
is Boltzmann suppressed away from the string core.
 Here $\alpha_{wk}$ is the SU(2) fine structure constant,
$M_W$ the mass of the W and
$\nu(T)$ the expectation value of the Higgs field. For this mechanism to work
baryon number violating processes
must occur in the core of the string at a greater rate than outside the string.
\p
The first step to finding the rate
of baryon number violating transitions is to calculate the height of the energy
barrier that separates the background states with different baryon number. If
the
barrier height is increased relative to the homogeneous background case, the
rate
of baryon number violation will probably be even more suppressed. On the other
hand,
if the barrier height is decreased, the exponential term in the rate will be
decreased,
but the prefactors must be calculated to obtain a firm result for the rate.
\p
In this paper we construct field space paths connecting purely bosonic string
background
states with different baryon numbers. The energy barrier obtained from each
path
provides an
upper bound to the barrier that would be obtained from the minimum action path.
\p
In section 2 we review the construction on the ordinary sphaleron in a form
that
can be
applied to string background fields. This method is then applied to various
string backgrounds:
superconducting string background in section 3 and  electroweak string
background in section 4. An
alternative for the superconducting string background is discussed in section 5
and we conclude in
section 6.  The calculations of  baryon number change and energy barriers are
presented in
appendices A and B respectively.
\vfil\eject
\noindent{\bf 2) The Ordinary Sphaleron}
\p
In this section we review the construction and properties of the electroweak
sphaleron [\klinkm].  The construction is presented in a form that is readily
generalised to string backgrounds and the various contributions to the
sphaleron
energy are highlighted.
\p
We take the bosonic part of the electroweak lagrangian to be
$$
{\cal L}=-{1\over 4}Y_{\mu\nu}Y^{\mu\nu}
-{1\over 4}W^a_{\mu\nu}W^{a\mu\nu} +(D_\mu\Phi)^\dagger
D^\mu\Phi-\lambda(\Phi^\dagger\Phi-{\nu^2\over 2})^2
$$
where  W is the
SU(2) triplet of gauge fields and Y is the U(1) gauge field. Using g and g'
to denote the  SU(2) and U(1) coupling constants,
the U(1) field strength is given by
$$
Y_{\mu\nu}=\partial_\mu Y_\nu -\partial_\nu Y_\mu,
$$
the SU(2) field strength by
$$
W^a_{\mu\nu}=\partial_\mu W^a_\nu -\partial_\nu W^a_\mu
-g\epsilon^{abc}W^b_\mu W^c_\nu,
$$
and the covariant derivative for the Higgs doublet $\Phi$
by
$$
D_\mu=\partial_\mu +{i\over 2}g\sigma^a W^a_\mu+{i\over 2}g'Y_\mu,
$$
where $\sigma^a$ are the Pauli matrices.
\p
In the case of vanishing Weinberg angle
the procedure for constructing a nontrivial path from
the ordinary vacuum, through the sphaleron back to the
ordinary vacuum is based on a unitary matrix, U. Using the notation of
ref.{\mog}
we obtain the nontrivial path from the trivial
vacuum configuration: $\Phi^\dagger= (0,v/\sqrt{2})$, $W=Z=A=0$,
by first considering the transformation
$$
\Phi \to \tilde\Phi =U\Phi\qquad W_\mu\to -{1\over g}(\partial_\mu U)
U^{-1}
$$
where
$$
U= \pmatrix{
\e^{i\mu}(\cos\mu -i\sin\mu\cos\theta) & \sin\mu\sin\theta\e^{i\phi}
\cr
-\sin\mu\sin\theta\e^{-i\phi} &\e^{-i\mu}(\cos\mu +i\sin\mu\cos\theta)
\cr},
$$
$\theta$ and $\phi$ are the usual spherical polar angles and $\mu$ is a
parameter.
U is an element of SU(2) and so this transformation would constitute a
gauge transformation if it could be applied everywhere. However, if
we simply apply the transformation everywhere,  the
Higgs field is not single valued on the polar axis and the gauge fields
are divergent.  To remove these problems profile functions are introduced:
$$
\Phi=(1-h(r))\pmatrix{0\cr {v\over \sqrt{2}}\e^{-i\mu}\cos\mu\cr}
+h(r)\tilde\Phi
$$
$$
 W_\mu = -{f(r)\over g}(\partial_\mu) U U^{-1}
$$
The boundary conditions on $h$ and $f$ are, $h,f \to 0$ as $r\to 0$ and
 $h,f \to 1$ as $r\to \infty$. The resulting field configuration is single
valued and finite. As the transformation differs from a gauge transformation
only when the profile functions differ from unity, the energy density of the
configuration is localised and the total energy is finite.
\p
The asymptotic Higgs field is given by  $\tilde\Phi$,
$$
\tilde\Phi = {v\over\sqrt{2}}\pmatrix{\sin\mu\sin\theta\e^{i\phi} \cr
e^{-i\mu}(\cos\mu +i\sin\mu\cos\theta) \cr}
$$
The nontrivial nature of this transformation is apparent if we consider the
winding of the asymptotic  Higgs field around the vacuum manifold.
We are interested in the Higgs field on some large shell surrounding the
sphaleron. Firstly the asymptotic Higgs field remains on the vacuum manifold,
$$
{lim \atop r\to\infty}\tilde\Phi^\dagger \tilde\Phi= v^2/2,
$$
and in general we can find one set of values for $\mu$, $\theta$ and $\phi$
that
correspond
to each point on the vacuum manifold. Thus the asymptotic Higgs field covers
the
vacuum manifold once during the transition from $\mu=0$ to $\mu=\pi$.
\p
The change in baryon plus lepton number during the transition can also be
calculated using the anomaly equation for general Weinberg angle,
$$
\partial_\mu J^\mu_{B+L} ={N_f \over 16\pi^2} (g^2W^{\mu\nu}_a
\tilde W_{a\mu\nu} -g'^2Y^{\mu\nu} \tilde Y_{\mu\nu})
$$
where $N_f$ is the number of families.
\p
Integrating over all space and the time interval of the transition  and
assuming
that we can drop the boundary terms, we have
$$
\Delta(B+L)=\int d^4x \bigl( {N_f \over 16\pi^2} (g^2W^{\mu\nu}_a
\tilde W_{a\mu\nu} -g'^2Y^{\mu\nu} \tilde Y_{\mu\nu}) \bigr)
$$
where $N_f$ is the number of families.
Explicit calculation with the field configurations given above yields
(see appendix A)
$$
\Delta(B+L)=2N_f
$$
Thus the transition not only covers the vacuum manifold but also produces
$N_f$
baryons.
\p
This process is greatly suppressed at low temperatures due to the energy
barrier
separating
the two vacua. The saddle point configuration
between the two vacua is called the sphaleron and is the field
configuration at $\mu=\pi/2$. The energy
density of the sphaleron configuration is (see appendix
B)
$$\eqalign{
{\cal E}_{\rm sphaleron}= &
{\nu^2\over 2}\hr^2 +{\nu^2\over r^2}h^2(1-f)^2
\cr &
+4(\fr)^2 {1\over g^2r^2} +{8\over r^4g^2}(1-f)^2f^2
+\lambda{\nu^4\over 4}(1-h^2)^2
\cr}
$$
This leads to a sphaleron energy of the form[\klinkm]
$$
{ E}_{\rm Sphaleron}={8\pi M_W\over g^2} D\bigl({\lambda\over g^2}\bigr)
$$
where D is a dimensionless factor.
Numerical integration yields[\klinkm]
D(0)=1.52, D(1)=2.07 and D($\infty$)=2.70.
\p
The large barrier between the vacua of different winding numbers leads to a
vast
suppression of baryon number
violating processes at low temperatures. At high temperatures the Boltzmann
suppression is reduced
and at very high temperatures the rate becomes O($\alpha T)^4$[\ambj].
\p
Strictly speaking the sphaleron is the field configuration that corresponds to
the saddle point separating
the two distinct vacua.  The sphaleron thus has the lowest possible
energy of any configuration that has the  maximum energy on a given path. In
the
general
setting we will refer to the highest energy configuration on a path as the
sphaleron although
we do not show that this is in fact the saddle point. In other words we will
construct upper bounds
on the energy barrier separating the background states.
\p
If we scale the radial variable in our expression for the sphaleron energy,
$r\to \eta x$, we find
$$\eqalign{
E_{\rm sphaleron} = 4\pi\eta\int dx   &\biggl[
{\nu^2\over 2} (x^2\hx^2 +2h^2(1-f)^2 )
\cr &
+4(\fx)^2 {1\over\eta^2 g^2} +{8\over \eta^2 x^2g^2}(1-f)^2f^2
\cr &
+{\lambda}{\nu^4\over 4}\eta^2 x^2(1-h^2)^2\biggr]
\cr}
$$
Extremising the energy with respect to the arbitrary scale factor allows us to
express the pure
gauge field contributions to the energy in terms of those involving the Higgs
field:
$$
\int dx\biggl[
{\nu^2\over 2} (x^2\hx^2 +2h^2(1-f)^2 ) +3{\lambda}{\nu^4\over 4}\eta^2
x^2(1-h^2)^2\biggr]
=\int dx\biggl[4(\fx)^2 {1\over\eta^2 g^2} +{8\over \eta^2
x^2g^2}(1-f)^2f^2\biggr]
$$
The sphaleron energy has two components, one coming purely from the gauge
fields
and the other
from the Higgs sector. As in the case of gauge topological defects, we can
think
of the scalar field
configuration being supported against collapse by the gauge field energy
trapped
in the core of the
object. With this picture in mind we see why sphaleron transitions in string
backgrounds might not
be as energetically disfavoured as those in a trivial background. In the core
of
a string the electroweak
symmetry is restored, the electroweak Higgs field is already forced to zero at
the core of the defect, so
at least this component of the sphaleron energy has already been 'paid for' by
the string. If there is a region of electroweak symmetry restoration that is
sufficiently large to accommodate the Higgs profile of
the sphaleron we might expect a reduction in the energy barrier by a factor of
about 0.5. If the region
 of symmetry restoration is much larger, as is the case for superconducting
cosmic strings, we might expect further reductions in the energy barrier as the
gauge fields can become more diffuse.

However, we must always bear in mind the cause of the symmetry restoration, in
the case of
superconducting strings the gauge field generated by the current not only gives
a large
region of symmetry restoration but also couples to the  sphaleron fields. This
situation is
discussed in the next section.

\vfil\eject
\noindent{\bf 3) Baryon Number Violating Transitions In A Superconducting
String
Background}
\p
In the next two sections we construct field space paths that connect string
states  with different baryon numbers.
The paths are constructed using the method outlined for the vacuum to vacuum
transitions. In each
case we consider the change in baryon number and, most
importantly, the energy barrier separating the initial and final states.  We
consider three cases, the
superconducting string, the electroweak string and the electroweak layer of a
nonsuperconducting
string formed at some transition above the electroweak scale. In each case we
apply the transformation
matrix U discussed above and introduce profile functions to ensure that the
fields are finite and
single valued everywhere.
\p
In this section we consider the superconducting string background. The
possibility of an enhanced
sphaleron transition rate in the region of electroweak symmetry restoration
around a superconducting
string was suggested by Damgaard and Esprin[\damg]. In contrast to the
background fields discussed
below, the background gauge field considered in ref.{\damg} was pure
hypercharge.
\p
Our first task is to construct the background state that will form the
endpoints
of our path.
In this section we consider the background electroweak fields provided by a
superconducting
GUT string.
A model of the electroweak fields around a GUT superconducting string is
discussed in detail
in ref.\pd, here we briefly review the main features of the field
configurations.
In this model the string carries a hypercharge current and so provides a
source for both the electromagnetic and the Z fields.
The electroweak fields around the string have the form:
$$
W^\pm_\mu=0 \qquad \Phi =\pmatrix{0\cr \phi(\rho)\cr} \qquad
Z_\mu=\delta_{\mu,z}Z(\rho)
\qquad A_\mu=\delta_{\mu,z}A(\rho)
$$
where $\rho$ is the  radial coordinate in a cylindrical polar coordinate
system.
The electromagnetic gauge field is free everywhere
outside the string and so has a logarithmic form out to scales where the string
curvature becomes
important.  The large Z field produced by the hypercharge current provides a
large, positive
contribution to the
Higgs field mass which leads to symmetry restoration in some region around the
string.  Inside the
symmetry restored region the Z field is massless and behaves logarithmically,
outside the region of
symmetry restoration the Z field becomes massive and decays exponentially.
For large string currents the profiles of the Higgs and Z fields can be
modelled
as
$$
\phi={1\over 2} (1+\tanh k(\rho-\rho_0)){\nu\over \sqrt{2}}
$$
$$
Z_3=\cases{{I\over 2\pi} \log({\rho\over \beta}) & $\rho<\rho_0$\cr
aK_0({1\over 2}G\nu \rho) & $\rho>\rho_0$\cr}
$$
and the three parameters  $r_0$, k and $\beta$ set to minimise the energy of
the
configuration. For
large currents this yields,
$$
 \rho_0={Ig\over 2\pi M_H M_W} ,\qquad
\beta \simeq\rho_0\e^{2/\nu G \rho_0}\simeq \rho_0
$$
The important features of these field configurations are the large region of
symmetry restoration and
the large background gauge field. For a maximal string current, $I\sim 10^{20}$
amps,
the region of symmetry restoration has a radius $r\sim 10^{10}$GeV$^{-1}\sim
10^{-5}$m.
The size of the symmetry restored region has been confirmed numerically
[\markt].
The form of the $W^3$ and hypercharge fields are important for the energetics
of
the baryon number
violating path. To find these we must also model the photon field around the
string. As the photon
field is free everywhere outside the GUT string, we take its profile to be
$$
A_z(\rho)= {\tilde I\over 2\pi }\log({\rho\over \gamma})
$$
We can find $\tilde I$ in terms of $I$ using the relationships between the Z
and
A fields
and the original SU(2) and hypercharge fields.

$$
\pmatrix{Z \cr A \cr}= \pmatrix{-\cos\theta_W & \sin\theta_W\cr
                                  \sin\theta_W & \cos\theta_W \cr}\pmatrix{W^3
\cr Y \cr}
$$
Thus in the symmetry restored region we have
$$
GW^3={-Ig\over 2\pi}\log({\rho\over \beta})+{\tilde I g'\over 2\pi
}\log({\rho\over \gamma}).
$$
The string does not act as source for $W^3$ and the field is free in the region
of symmetry
restoration, thus $W^3$ should be constant in this region. This leads to
$$
A_z(\rho)= { gI\over 2\pi g'}\log({\rho\over \gamma}).
$$
The scale in the logarithm cannot be determined
for an infinitely long, straight string. However, if we imagine the string to
be
curved we can
determine a value for $\gamma$. In particular, if we consider the string to
form
a circular loop of
radius $R$, the natural value for $\gamma$ is of order $R$[\menpb]. As the
typical curvature scales are
of order kiloparsecs, this curvature has no further appreciable affect. The
forms for the  SU(2)
and hypercharge fields are then,
$$
Y_z= \cases{
{I\over 2\pi}{1\over Gg'} \bigl(g'^2\log({\rho\over \beta}) +g^2\log({\rho\over
\gamma})\bigr)
   & $\rho<\rho_0$\cr
{I\over 2\pi}{g^2\over Gg'}\log({\rho\over \gamma}) +{g' \over G}aK_0({1\over
2}G\nu \rho)
 & $\rho>\rho_0$\cr}
$$
$$
W^3_z= \cases{
{I\over 2\pi}{g\over G} \log({\beta\over \gamma})
   & $\rho<\rho_0$\cr
{I\over 2\pi}{g\over G} \log({\rho\over \gamma}) -{g \over G}aK_0({1\over 2}
G\nu
\rho)
 & $\rho>\rho_0$\cr}
$$
Thus $W^3_z$ is large and constant within the symmetry restored region.
\p
The most naive generalisation of the procedure outlined in section 2 for the
production of baryon
number violating paths is to apply the same transformation U as in the vacuum
case. As we will see,
this produces a transformation with the required change in baryon number,
however we have no
guarantee that this is the lowest energy path between the string background
states. Indeed, given
the different symmetries of the string and ordinary sphaleron it seems unlikely
that this procedure
will generate the lowest energy path, even for zero Weinberg angle.
 We will return to this point once we have calculated the energy
of the {\it sphaleron} derived from this naive approach.
\p
Applying the transformation discussed in section 2 to the asymptotic Higgs
field
we find,
$$
\Phi\to U\Phi= {\nu\over \sqrt{2}}\pmatrix{\sin\mu\sin\theta \e^{i\phi} \cr
\e^{-i\mu}(\cos\mu+i\sin\mu\cos\theta)\cr} p
$$
where $  p$ is the profile function for the Higgs field which satisfies the
boundary conditions $p\to 0$
as $\rho\to 0$ and $p\to 1$ as $\rho\to \infty$. Apart
from the region close to the string core the Higgs field lies on its vacuum
manifold and we can show,
as in the
case of the ordinary sphaleron, that
every point on the vacuum manifold is covered at some
stage  during the transition by the Higgs field at some point on any large
shell
around the sphaleron.
Possible complications arise as the string will always pierce this large shell
along the polar axis. The
fact that the Higgs field vanishes at both $\theta=0$ and $\theta=\pi$ means
that the lines on the
vacuum manifold that were covered by the north and south poles are no longer
covered. The north pole,
$\cos\theta=1$, is invariant during the transition while the south pole,
$\cos\theta=-1$ is mapped to
$$
{\nu\over \sqrt{2}}\pmatrix{0\cr \e^{-2i\mu}\cr}.
$$
Apart from this line the whole vacuum manifold is covered.

\p
We can verify that this transition violates baryon number by explicitly
calculating
$\Delta(B+L)$. The gauge fields in this case are given by
$$
\vec W_\mu \to U \tau^a W^a_\mu U^{-1} -{f\over g}(\partial_\mu U)U^{-1}
=U\delta_{\mu,z}\tau^3 W^{3{(\rm string)}}_zU^{-1} -{f\over g}(\partial_\mu
U)U^{-1}
$$
where we have again introduced a profile function, $f$. $W^{3({\rm string})}_z$
is the gauge field generated by the current in the string, it satisfies the
boundary conditions,
$W^{3({\rm string})}_z(\rho=0)={\rm const.}$ and
$W^{3({\rm string})}_z(\rho\to\infty)\to{I\over 2\pi}{g\over G}
\log({\rho\over \gamma}) $.
In this case the cylindrical symmetry of the string
and the spherical nature of the transformation suggest that we let $f$
 depend on both $\theta$ and r
(we choose to work in spherical polars). Similarly for $\mu\neq 0$ we allow the
Higgs field profile
function $p$
to depend on both $\theta$ and r. Generalising the profiles in this way is our
only concession to
the mixed nature of the symmetries of the problem.
\p
We can explicitly calculate the change in baryon plus lepton number that the
transformation
induces by substituting the gauge field configurations into the anomaly
equation
(see App.A).
We find
$$
Tr(W_{tr}W_{\theta\phi}-W_{t\theta}W_{r\phi}+W_{t\phi}W_{r\theta})
={\rm pure \hskip 6pt sphaleron \hskip 6pt contribution}
+\sin(\mu)\cos(\mu)(\mu\hskip 6pt
{\rm indep.})
$$
where the pure sphaleron contribution is as discussed in the previous section;
$$
12(\cos(\theta)-1)\sin^2(\mu)\fr{f(1-f)\over r^2g^2}
$$
Integrating from $\mu=0$ to $\mu=\pi$
removes the contribution from the string gauge field, thus the string gauge
field
does not alter the change in baryon number caused by the transition. We now
have
a path that interpolates
between string background states of different baryon number.
\p
The energy density associated with the static field configurations at any point
on this path
is calculated in App.B. It is convenient to work with
$W(r,\theta)=gW^3_z(r,\theta)$ and
$Y(r,\theta)=g'Y_z(r,\theta)$ when calculating
the energy of the sphaleron. The boundary conditions on $W$ are then simply
$W\to {\rm constant}$ at the string core
and $W\sim \log(r\sin\theta/\gamma)$ far from the string. The energy density is
found to be:
$$\eqalign{ {\cal E}= &
4\fr^2{\sin^2\mu \over g^2r^2} +{8\over r^4g^2}\sin^4\mu(1-f)^2 f^2
 +{2\over g^2 r^4} \sin\mu^2 \fth^2
\cr &
+{1\over 2g^2r^2}(\sin\theta r\Wr+\cos\theta\Wth)^2
\cr &
+{2\over g^2r^2} \fr\sin\mu \cos\mu\sin\theta(\sin\theta r\Wr+\cos\theta\Wth)
\cr &
+8W{ \sin^3\mu\over g^2 r^3} \sin^2\theta\cos\mu  f (1-f)^2
\cr &
+2 {W^2 \over g^2r^2} \sin^2\mu (1-f)^2 (\cos^2\mu\sin^4\theta +2\cos^2\theta)
\cr &
+{1\over 2g'^2r^2}(\sin\theta r\Yr+\cos\theta\Yth)^2
\cr &
+{\nu^2\over 2}\hr^2 +{\nu^2\over 2r^2}\hth^2 +{\lambda \nu^4\over 4}(1-h^2)^2
\cr &
+{\nu^2h^2\over 2} \bigl({1\over 4}(W-Y)^2
+{2\over r^2} \sin^2\mu (1-f)^2-{1\over r}\sin\mu\cos\mu (W-Y) (1-f)
\sin^2\theta\bigr)
\cr}
$$
We can rewrite some of these terms as a sum of squares,
$$
\eqalign{
& {8\over r^4g^2}\sin^4\mu(1-f)^2 f^2
+8W{ \sin^3\mu\over g^2 r^3} \sin^2\theta\cos\mu  f (1-f)^2
\cr &
+2 {W^2 \over g^2r^2} \sin^2\mu (1-f)^2 (\cos^2\mu\sin^4\theta +2\cos^2\theta)
\cr &
={2\over r^4g^2}\sin^2\mu(1-f)^2\bigl(
[2f\sin\mu+rW\sin^2\theta\cos\mu]^2+2r^2W^2\cos^2\theta\bigr)
\cr}
$$
Thus in the string core where we have the background  fields $h=0$,  W$\sim$
constant,
the energy density is manifestly a sum
of positive definite terms and we can place a lower bound on the energy
density,
$$ {\cal E}>
4\fr^2{\sin^2\mu \over g^2r^2}
+{2\over r^2g^2}\sin^2\mu(1-f)^2\bigl(
2W^2\cos^2\theta\bigr)
$$
We can place a lower bound on the energy of the configuration by considering
the
integral of the
energy density in regions within $45^o$ of the poles, i.e. regions with
$\cos^2\theta>1/2$.
A lower bound to this polar contribution to the energy is then obtained by
setting $\cos^2\theta=1/2$,
leading to
$$
E_{\rm polar}> [2\pi][2(1-1/\sqrt{2})]
\int r^2dr \biggl(4\fr^2{\sin^2\mu \over g^2r^2}
+{2\over r^2g^2}\sin^2\mu(1-f)^2 W^2\biggr)
$$
Recalling that  W is a constant inside the region of symmetry restoration and
imposing the boundary conditions $f\to 0$ as $r\to 0$ and $f\to 1$ as $r\to
\infty$, the integral is extremised by taking $f=1-\exp(-\vert W\vert
r/\sqrt{2})$ and then has the value
$2\sqrt{2}\sin^2\mu\vert W\vert/g^2$. This gives the bound
$$
E_{\rm polar}> 4\pi(\sqrt{2}-1)\sin^2\mu {I\over \pi}{1\over G}\log({\gamma
\over
\beta})
\sim E_{\rm Sphal} \sin^2\mu (M_H/{\rm GeV})(I/10^{10}{\rm
amps})\log({\gamma\over \beta})
$$
Thus there is a lower limit to the sphaleron energy determined by the string
current and  the sphaleron
energy is increased at large string currents. The origin of this increase is
indicated by the above
analysis: in the core of the sphaleron the transformation is not pure gauge and
so physical gauge
fields are created. In the string background these gauge fields acquire a mass
through the nonabelian
terms in the SU(2) field strength which couple them to the Z field. Exciting
these fields then entails
an energy cost of order the induced mass.
\p
 Although this result has been determined for a specific
sphaleron configuration, the above discussion suggests that a similar result
will hold for any
configuration that excites massive gauge fields. With the above form for the
background field,
$W^\pm_t$, $W^\pm_x$
and  $W^\pm_y$ get masses of order I.  Working temporarily in cartesian
coordinates, the change in
baryon number can be rewritten as,
$$
\int d^4x W^{a\mu\nu}\tilde W_{\mu\nu}^a =\int d^4x \partial_\mu\biggl(
\epsilon^{\mu\nu\lambda\sigma}(W^a_{\nu\lambda} W^a_\sigma -{2\over 3}
g\epsilon^{abc}
W^a_\nu W^b_\lambda W^c_\sigma)
\biggr)
$$
If we set the massive gauge fields to zero the second term above vanishes.  We
can now
use the divergence theorem to transform the quantity of interest into a surface
integral.
The integrand contains a factor of $W^a_{\nu\lambda}$ and so vanishes except
where the
string cuts the surface. The nonvanishing components of the field strength for
the string
are $W^3_{13}$ and $W^3_{23}$. Any transformation that produces further
components
asymptotically will have infinite energy and can be neglected. For a
hypercubical surface
the faces of interest are $t=t_i$, $t=t_f$ and $z\to -\infty$, $z\to +\infty$.
The starting
and finishing configurations should be identical, so the contributions from the
$t=t_i$, $t=t_f$ planes should cancel. The contribution from the $z\to
-\infty$,
 $z\to +\infty$
planes is removed by the Levi-Civita tensor. The net change in B+L is thus
zero.
\p
Alternatively we work directly with the original form.
With the massive gauge fields set to zero we  have $W^A_{xy}=0$, $W^A_{ty}=0$
and $W^A_{tx}=0$
where A=1,2 and the quantity of interest reduces to
$$
W^{a\mu\nu}\tilde W_{\mu\nu}^a
={4}(W^3_{tx}W^3_{yz}-W^3_{ty}W^3_{xz}+W^3_{tz}W^3_{xy})
$$
$$
={4}([W^3_{x,t}-W^3_{t,x}][W^3_{z,y}-W^3_{y,z}]
-[W^3_{y,t}-W^3_{t,y}][W^3_{z,x}-W^3_{x,z}]+[W^3_{z,t}-W^3_{t,z}][W^3_{y,x}-
W^3_{x,y}])
$$

The quantity we are interested in is the integral of the above over all space
and the time
interval of the transition. Integration by parts allows us
to exchange the derivatives in each term at the expense of some boundary terms.
This takes two
integrations by parts and so preserves the sign of each term, however the
alternating nature of
the Levi-Civita tensor means that the object is odd under interchange of two
indices, hence
the integrand acquires an overall factor of -1. Some of the boundary terms
contain factors of
the form $W^3_{xy}$, if the energy of the configuration is finite these must
vanish asymptotically.
A further integration by parts transforms the remaining terms to this form and
they vanish
likewise. Thus there are no finite energy baryon number changing paths that do
not excite massive
gauge fields and we expect the increased barrier height found in the specific
example to be a
generic feature of the superconducting string background.
\p
We conclude from this argument that the energy barrier separating
superconducting string states of
different baryon number is large for transitions that occur within the region
of
electroweak symmetry restoration.  This increase in barrier height occurs
despite the restoration
of the electroweak symmetry and is due to the presence of the large background
gauge field. In this case
the Higgs sector contributes very little to the energy barrier whilst the pure
gauge field
contributions are greatly increased.
\p
In the following sections we investigate situations in which the restoration of
electroweak symmetry is
not a result of a large background gauge field.
\vfil\eject
\noindent{\bf 4) The Electroweak String Background}
\p
There are two mechanisms which can cause the restoration of the electroweak
symmetry below the
critical temperature, the first is via an interaction with some background
field, the superconducting
string is an example of this, alternatively the electroweak Higgs field may
{\it
wind} and be forced
to zero in some region for reasons of continuity, this is the case with
electroweak strings. Although
the standard model does not admit topologically stable string solutions it has
been shown[\nambu] that
string solutions do exist and that they are energetically stable for some
parameter values[\mogtan].
These strings
take the form of Nielsen-Olesen vortices[\vort] formed from the lower component
of the Higgs doublet and the
Z field, with all other fields set to zero. The nonvanishing fields are (in
cylindrical polar coordinates ($\rho,\phi)$),
$$
\Phi= \pmatrix{0\cr p\e^{-i\phi}\cr} \quad Z_\phi= \tilde Z/\rho
$$
where for small $\rho$: $p\propto \rho$ and $\tilde Z\propto \rho^2$ while at
large $\rho$: $p\to \nu/\sqrt{2}$
and $\tilde Z \to 2/G$ ($G^2=g^2+g'^2$).
\p
Although electroweak strings are stable for some parameter values,  they are
unstable
for the observed parameter values. This complicates the
interpretation of the energies we will calculate for parameter values outside
the stability region. In the case of the superconducting string we had a
stable background field configuration and we could take the barrier height to
be
the difference in
energy between the pure string configuration and the string plus sphaleron
configuration.  In the
electroweak string case we don't always have a stable background configuration
and so
in these
cases we cannot simply  measure
the energy difference between the pure string and string plus sphaleron states.
For these
parameter values we consider
the electroweak string background as an example of electroweak symmetry
restoration due to the presence
of strings from some higher energy phase transition (for example the
technicolour model discussed in
 ref.\bdt). The presence of the higher energy string leads to electroweak
symmetry restoration in its
vicinity. This pinning of the electroweak Higgs field ameliorates the prime
instability of the
electroweak string by preventing the upper component of the Higgs field
acquiring an expectation
value in the core of the string. Whether this is sufficient to stabilise the
electroweak string
configuration for physical parameter values
remains to be investigated. We will assume that the presence of the higher
energy
string stabilises the electroweak string configuration for parameter values
outside the
usual stability region. (Alternatively we could imagine the electroweak
Higgs field simply pinned to zero in the core of the higher energy string. The
scale of symmetry
restoration in this case depends on the details of the interaction between the
fields. We can't ignore
the finite size of the higher energy string as if we simply
treat the pinning as a boundary condition at $r=0$, Derricks theorem tells us
that the Higgs field
expectation value should rise arbitrarily rapidly to its vacuum value.)
\p
Once again we apply the transformation discussed in section 2, allow the
profile
functions to depend on $r$
and $\theta$ and
evaluate the change in baryon number and the energy density of the static field
configurations. From App.A
we see that the extra contributions to the change in baryon number from the
string gauge field vanish when we integrate over all space:  once again we have
constructed a baryon
number changing path.
\p
  From App.B we see that the energy density is,
$$\eqalign{ {\cal E} = &
{1\over 2g^2 r^2} \Sr^2 +{1\over 2g^2 r^4} (\Sth +S\cot\theta)^2
\cr &
+{1\over 2g'^2r^2} \Yr^2 +{1\over 2g'^2r^4}(\Yth+Y\cot\theta)^2
\cr &
\bigl[\hr^2 +{1\over r^2}\hth^2 +{h^2\over  r^2} \bigl({S\over 2}-{Y\over 2}
+{1\over \sin\theta}\bigr)^2 \bigr]{\nu^2\over 2}
\cr &
+{\lambda \nu^4\over 4}(1-h^2)^2
\cr +\sin^2\mu \biggl[ &
4 \fr^2 {1\over g^2 r^2}+2 \fth^2 {1\over g^2 r^4}  +{2\sin\theta\over g^2 r^2}
[\fr \Sr +{1\over r^2}\fth\Sth]
\cr &
+(6 -4f) \fth  S {\cos\theta \over g^2 r^4}  -4 \Sth (1-f) f {\cos\theta \over
g^2 r^4}
\cr &
-{4\cos^2\theta\over g^2 r^4 \sin\theta} S f (1-f)
 +2 S^2 {\cos^2\theta\over g^2 r^4}  (1-f)^2
\cr &
- {h^2\nu^2\over 2r^2}[2 f+\sin\theta(S-Y)] (1-f)
 \biggr]
\cr +\sin^4\mu &
{2\over r^4 g^2} [2f+S\sin\theta]^2 (1-f)^2
\cr}$$
where $S=grW^3_\phi$ and $Y=g'rY_\phi$
Rewriting the gauge fields in terms of the photon and Z fields, $G A= g' W^3+g
Y$ and $G Z=g' Y-g W^3$,
asymptotically we have only kinetic terms for $ A$. Thus we have the usual
forms
for the electroweak
string gauge fields at large distances from the string core, $A=0$ and $Z_\phi
=2/(Gr\sin\theta)$.
In terms of the original SU(2) and hypercharge fields the asymptotic forms are,
$$
W^3_\phi=-{g\over G}{2\over Gr\sin\theta},\quad Y_\phi={g'\over G}{2\over
Gr\sin\theta},
$$
or equivalently,
$$
S=-{g^2\over G^2}{2\over\sin\theta},\quad Y={g'^2\over G^2}{2\over\sin\theta}.
$$
\p
The expression for the energy density simplifies if we define
$$
S={T\over \sin\theta},\quad Y={V\over \sin\theta}.
$$
Working in cylindrical polar coordinates with the scaled variables
$k\rho=r\sin\theta$ and
$kz=r\cos\theta$,
the energy density then  becomes,

$$\eqalign{ g^2k^4{\cal E} = &
{1\over 2 \rho^2} (T,_\rho^2 +T,_z^2)
+{g^2\over 2g'^2\rho^2}( V,_\rho^2 +V,_z^2)
\cr &
\bigl[\bigl(h,_\rho^2 +h,_z^2\bigr)
+{h^2\over  \rho^2} \bigl({T\over 2}-{V\over 2} +1\bigr)^2
\bigr]{g^2k^2\nu^2\over 2}
+{\lambda \nu^4g^2k^4\over 4}(1-h^2)^2
\cr +\sin^2\mu \biggl[ &
{2\over r^2}(f,_\rho^2 +f,_z^2)+{2\over r^2}[\cos^2\theta f,_z^2
+2\cos\theta\sin\theta
f,_zf,_\rho +\sin^2\theta  f,_\rho^2] +{2\over r^2} [f,_\rho T,_\rho +f,_z
T,_z]
\cr &
+4(1 -f) {\cot\theta \over r^3} [ T(-\sin\theta f,_z+\cos\theta f,_\rho)
                                 -f(-\sin\theta T,_z+\cos\theta T,_\rho )]
\cr &
+2 T^2 {\cos^2\theta\over \sin^2\theta}{1\over r^4}  (1-f)^2
- {h^2\nu^2g^2k^2\over 2 r^2}[2 f+T-V] (1-f)
 \biggr]
\cr +\sin^4\mu &
{2\over r^4} [2f+T]^2 (1-f)^2
\cr}$$
\p
Numerical minimisation of this integral gives an energy shift relative to the
pure string background
of order 50\% of the ordinary sphaleron energy if we set $\sin\theta_W$ close
to
one  (in this case
the electroweak string is stable). We might expect such a reduction in the
energy barrier as
the energy
contribution from the Higgs sector will be small since the region of
electroweak
symmetry restoration
around the string is of an appropriate size to accommodate the Higgs profile of
the sphaleron. If we
set $\sin\theta_W$ to its observed value, the energy shift relative to the
string background
at $\sin\mu=\pi/2$ is
of order 10\% of the ordinary sphaleron energy.
As we saw in section 2, the Higgs sector contributes less than half of the
energy of the sphaleron, thus
there is another effect at work in this case.
This second effect is the interaction of the string gauge fields with
the sphaleron gauge fields. Recalling that T is negative while $f$ and V are
positive we see that there
is the potential for partial cancellation of several terms leading to a further
reduction in the sphaleron
energy.  This observation raises the question of interpretation: one mode of
instability in the
electroweak string is the formation of a W condensate in the string
core[\meprd],
is this being excited here?
The gauge fields have the appropriate angular dependence, so we are once again
faced with the question
of where to measure the barrier height from.
\p
We have constructed a baryon number violating transition in the background of
an
electroweak string. Once
again the gauge fields of the string play an important role in determining the
energy barrier between
states of different baryon number.  In this case there may be a considerable
decrease in the sphaleron
energy, but a detailed analysis of the couplings between the electroweak sector
and the higher energy
sector producing the strings is required before a definite conclusion can be
drawn.
\vfil\eject
\noindent{\bf 5) An Alternative Transformation In the Superconducting String
Background}
\p
The paths we have considered so far are based on the matrix $U$ discussed in
section 2. This matrix was constructed so as to produce a spherically symmetric
 sphaleron.  When we introduce strings the natural symmetry is cylindrical
rather
than spherical and we might expect the sphaleron to deform. For example we
might consider  the sphaleron becoming prolate. As an extreme case we
can take the transformation on the equatorial plane of  sphaleron and apply
this
along the whole length of the string i.e.  we take $U$ and set $\theta=\pi/2$,
giving
the transformation
$$
U'=\pmatrix{
\e^{i\mu}\cos\mu  & \sin\mu\e^{i\phi}
\cr
-\sin\mu\e^{-i\phi} &\e^{-i\mu}\cos\mu
\cr}.
$$
Applying this transformation to the superconducting string background,
we find that the change in B+L is given by
$$\eqalign{
c\Delta (B+L)=\int d^4x\quad 2\dot{\mu}{\sin\mu\cos\mu \over g^2
r^3\sin\theta}\biggl(  &
{\partial\over \partial\theta}(S\cos\theta(f-1)^2) +(f-1){\partial\over
\partial\theta}(S\cos\theta) \cr &
+r\sin\theta[{\partial\over \partial r}(S(f-1)^2) +(f-1){\partial\over \partial
r}S]\biggr)
\cr}$$
where $S=grW^3_z$ and $c=-2\pi^2/N_f g^2$.
The string gauge field vanishes at the centre of the string, so the first term
vanishes when integrated
with respect to $\theta$.  After integrating the second term by parts with
respect to $\theta$ and the
remaining terms with respect to $r$ we have
$$\eqalign{
c\Delta (B+L)=4\pi\int dt \dot{\mu}{\sin\mu\cos\mu \over g^2 } \biggl( &\int
drd\theta
(-S\cos\theta) {1\over r}{\partial f\over \partial\theta} \cr &
+\int d\theta\sin\theta [S(f-1)^2 +(f-1)S]_0^\infty \cr &
-\int drd\theta \sin\theta S{\partial f\over \partial r}\biggr)
\cr}$$
Now, $\cos\theta {1\over r}{\partial f\over \partial\theta}
+\sin\theta {\partial f\over \partial r} ={\partial f\over \partial \rho}
$ where $\rho$ is the radial coordinate in cylindrical polars.
$$
c\Delta (B+L)=4\pi\int dt \dot{\mu}{\sin\mu\cos\mu \over g^2 } \bigl( \int
drd\theta
(-S{\partial f\over \partial\rho})
+\int d\theta\sin\theta [S(f-1)^2 +(f-1)S]_0^\infty \bigr)
$$
As S vanishes at $r=0$, the lower limit of the boundary term gives zero. Far
from the string
$f$ tends to one, so the only contribution from the upper limit comes from the
string core.
$ [S(f-1)^2 +(f-1)S]$ is finite and only nonzero in regions with $\sin\theta
<\rho_c/R$  where
$\rho_c$ is the core radius and $R$ is the spherical distance we are
considering. As  $R\to\infty$
the contribution from the string core vanishes and we are left with
$$
c\Delta (B+L)=4\pi\int dt \dot{\mu}{\sin\mu\cos\mu \over g^2 }\int drd\theta
\bigl(-S{\partial f\over \partial\rho} \bigr)
$$
S has a fixed sign and $f$ is monotonically increasing away from the string
core, thus the
integrand has a fixed sign and the spatial integrals are nonzero.
The temporal integral gives a factor of $[-\cos(2\mu)/2]_i^f$, thus the vacuum
($\mu_i=0$) to
vacuum ($\mu_f=\pi$) transition gives no net change in B+L. However, there is a
change in
B+L as we move from the vacuum ($\mu_i=0$) to $\mu=\pi/2$.  The field
configurations at $\mu=\pi/2$
are
$$
\tilde\Phi = {v\over\sqrt{2}}\pmatrix{\e^{i\phi} \cr
0 \cr} p(\rho)
$$
$$
 W_\mu =U' \delta_{\mu,z} \tau^3 {S\over gr}U'^{-1}
 -{f\over g}\partial_\mu U' U'^{-1}
=-\delta_{\mu,z} \tau^3 {S\over gr}-{f\over g}\delta_{\phi\mu}\pmatrix{i &0\cr
0
&-i\cr}
$$
Thus we have the superconducting string
superimposed on an electroweak Higgs field with a winding number, nonvanishing
$W^3_z$ and $W^3_\theta$
but vanishing $W^\pm_\mu$.
\p
The energy density for static configurations generated by this
transformation is
$$\eqalign{
{\cal E}= &{1\over 2\sin^2\theta}{1\over r^4g^2}
\biggl(4\sin^2\mu(\fth^2+r^2\fr^2)+r^2\sin^4\theta\Sr^2  \cr &
            -2\Sr rS\sin^4\theta
             +\Sth^2\cos^2\theta\sin^2\theta +2\cos\theta\sin^3\theta\Sth(r\Sr
-S) \cr &
             +S^2[\sin^4\theta +4\cos^2\mu(1-f)^2\sin^2\mu]   \biggr)  \cr &
              +{\nu^2\over 2}\biggl[\hr^2 +{1\over r^2}\hth^2 +{h^2\over
r^2}{1\over \sin^2\theta}(1-f)^2\sin^2\mu +{h^2\over 4r^2}(S-Y)^2\biggr] \cr &
             +{\lambda\nu^4\over 4}(1-h^2)^2
             +{1\over 2r^4g'^2}(\Yth\cos\theta+(\Yr-Y/r)\sin\theta r)^2
\cr}$$
where $Y=g'rY_z$.
The interesting feature in this case is the form of the $S^2$ term. The
contribution due to the
transformation has the form $S^2(1-f)^2\cos^2\mu\sin^2\mu$, thus it vanishes at
$\mu=\pi/2$.
This is what we would expect from our discussion in section 3 given the absence
of $W^\pm$ fields in this
case. We thus expect to incur a large energy cost if we try to deform away from
this configuration.
Although a detailed stability analysis would be required to prove the stability
of the electroweak sector
with this winding, this observation does suggest that the winding might be
stable. The background gauge
field generated by the current reduces the two main modes of instability of the
electroweak string.
Given that electroweak strings are themselves superconducting, it would be
interesting if the
electroweak string could carry its own stabilising current.
\p
In the background of a superconducting string it is possible that windings in
the electroweak Higgs field
are stabilised and such configurations carry a net baryon plus lepton number.
\vfil\eject
\noindent{\bf Conclusions}
\bigskip
We have constructed baryon number changing transitions in various string
backgrounds. The energy
barriers in the various backgrounds depend crucially on the region of symmetry
restoration and
the background gauge fields. For a simple dip in the vacuum expectation value
of
the electroweak
Higgs field, generated for instance by a scalar coupling to a higher energy
string, the sphaleron
barrier can be reduced to about half its standard size although the precise
reduction in the barrier height depends on the details of the model. In the
background of an electroweak string the gauge fields can be exploited to
further
lower the sphaleron
barrier, unfortunately the interpretation of this result is complicated by
considerations of the
stability of the string.
However, in the case of the superconducting cosmic string
background the gauge fields that produce the large region of electroweak
symmetry restoration also
generate masses for the $W^\pm$ fields, increasing the height of the sphaleron
energy barrier.

It is possible to reduce the sphaleron energy barrier in the background of a
cosmic string, but the
scale of the reduction depends on the details of the model and background gauge
fields can even
increase the barrier height. This causes problems for the baryogenesis
mechanism
of ref.{\bdt}
 that is based on superconducting cosmic strings. The mechanism based on
technistrings may be viable, but a
detailed model of the electroweak background fields is required.

In the presence of a superconducting cosmic string windings of the electroweak
Higgs field may be
stabilised and such configurations carry baryon number.
\bigskip
\noindent{\bf Acknowledgements}
\bigskip
I would like to acknowledge A.C.Davis for interesting me in this project and
N.Dorey and S.Hands for
useful discussions.
\bigskip
\noindent{\bf References}
\def\jump{\hskip 0.05truein}
\medskip
\noindent{[}1] N.Turok and T.Zadrozny Phys.Rev.Lett{\bf 65} (1990) 2331

\jump L.McLerran, M.Shaposhnikov, N.Turok and M.Voloshin Phys.Lett.{\bf B256}
(1991) 451

\jump A.Cohen, D.Kaplan and A.Nelson Phys.Lett.{\bf B263} (1991) 86

\jump A.Nelson,  D.Kaplan and A.Cohen Nucl.Phys.{\bf B373} (1992) 453

\noindent{[}2] M.Dine, R.Leigh, P.Huet, A.Linde and D.Linde Phys.Rev.{\bf D46 }
(1992) 550

\noindent{[}3] R.Brandenberger, A.C.Davis and M.Trodden Phys.Lett.{\bf 335 }
(1994) 123

\noindent{[}4] P.Arnold and L.McLerran  Phys.Rev.{\bf D36 } (1987) 581, Phys.
Rev. {\bf D37 } (1988) 1020

\noindent{[}5] F.R.Klinkhamer and N.Manton Phys.Rev.{\bf D30 } (1984) 2212

\noindent{[}6] M.James Phd Thesis DAMTP Cambridge (1993)

\noindent{[}7] J.Ambj\o rn, M.Laursen and M.Shaposhnikov Phys.Lett.{\bf B197}
(1987) 49

\jump J.Ambj\o rn, T.Askgaard, H.Porter and M.Shaposhnikov Nucl.Phys.{\bf B353}
(1991) 346

\noindent{[}8] P.Damgaard and D.Espriu Phys.Lett.{\bf B256} (1991) 442

\noindent{[}9] W.B.Perkins and A.C.Davis Nucl.Phys.{\bf B406 } (1993) 377

\noindent{[}10] M.Trodden Mod. Phys.Lett.{\bf A9} (1994) 2649

\noindent{[}11] W.B.Perkins Nucl.Phys.{\bf B364} (1991) 451

\noindent{[}12] T.Vachaspati Phys.Rev.Lett.{\bf 68} (1992) 1977

\jump Y.Nambu Nucl.Phys.{\bf B130} (1977) 505

\noindent{[}13] M.James, L.Perivolaropoulos and T.Vachaspati  Nucl.Phys.{\bf
B395} (1993) 534

\noindent{[}14] H.B.Nielsen and P.Olesen Nucl.Phys.{\bf B61 } (1973) 45

\noindent{[}15] W.B.Perkins Phys.Rev.{\bf D47 } (1993) 5224
\vfil\eject

\noindent{\bf Appendix A. Calculating the Change in B+L}
\p
\noindent{\bf The Ordinary Sphaleron}
\p
The quantity that we need to calculate is $\int d^4x W\tilde W$
where $\tilde W$ is the dual of W, $\tilde W^{\mu\nu}= {1\over
2}\epsilon^{\mu\nu\sigma\rho}
W_{\sigma\rho}$.  Using the symmetry of the
alternating tensor and of the field strength, we have three
distinct contributions to calculate:
$$
W_{tr}W_{\theta\phi} \quad , \quad W_{t\theta}W_{r\phi}
\quad {\rm and}\quad W_{t\phi}W_{r\theta}
$$
Using the pure sphaleron gauge fields
$$
W_\mu  =-{f(r)\over g}(\partial_\mu U) U^{-1}
$$
with the aid of Maple we find
$$
Tr(W_{tr}W_{\theta\phi} -W_{t\theta}W_{r\phi} + W_{t\phi}W_{r\theta})=
12{\partial f\over \partial r}f \sin^2(\mu)(1-\cos(\theta)){f-1\over g^2 r^2}
$$
Integrating over space we have
$$
\int_0^{2\pi} d\phi\int_0^\pi \sin\theta d\theta \int_0^\infty r^2 dr
Tr(W_{tr}W_{\theta\phi}
-W_{t\theta}W_{r\phi} + W_{t\phi}W_{r\theta})
$$
$$
={12\sin^2(\mu)\over g^2}
\int_0^{2\pi} d\phi\int_0^\pi \sin\theta d\theta (1-\cos(\theta))\int_0^\infty
dr{\partial f\over \partial r}f
(f-1)
$$
$$
={12\sin^2(\mu)\over g^2}[2\pi] [2][{f^3\over 3}-{f^2\over 2}]_0^\infty
=-{8\pi\over g^2}\sin^2\mu
$$
where we have used the boundary conditions on $f$,  $f(0)=0$ and $f(\infty)=1$.
Thus we have
$$
\Delta(B+L)={N_f\over 16 \pi^2}\int d^4x g^2W^{\mu\nu}_a \tilde W_{a\mu\nu}
={N_f\over 16 \pi^2}\int d^4x g^2{1\over 2}
\epsilon^{\mu\nu\lambda\rho}W_{a\mu\nu}
  W_{a\lambda\rho}
$$
$$
= {4N_f\over \pi}\int_0^\pi d\mu\sin^2(\mu)= 2N_f
$$
where we have used the normalisation condition,
$
Tr(t^a t^b) =-{1\over 2}\delta^{ab}
$.
\p
Now that we have calculated the pure sphaleron contribution to the change in
$B+L$ we can
see what changes the string backgrounds make.
\bigskip
\noindent{\bf The Superconducting String Background}
\p
In the case of superconducting strings we have the gauge field discussed in
section 3,
$$
\vec W_\mu=U\delta_{\mu,z}\tau^3 {W(r,\theta)\over g}
U^{-1} -{f(r,\theta)\over g}(\partial_\mu U)U^{-1}
$$
where $W$ is a constant inside the region of symmetry restoration.
 Substituting these fields into
$Tr(W_{tr}W_{\theta\phi}-W_{t\theta}W_{r\phi}+W_{t\phi}W_{r\theta})$
we find
$$\eqalign{
Tr(W_{tr}W_{\theta\phi}-W_{t\theta}W_{r\phi}+W_{t\phi}W_{r\theta})
 =&12(\cos(\theta)-1)\sin^2(\mu)\fr{f(1-f)\over r^2g^2}
\cr &
+\sin(\mu)\cos(\mu)(\mu\quad{\rm indep.})\cr}.
$$
The term that is independent of $\mu$ carries no explicit $\mu$ dependence, but
does contain
various profile functions. We can calculate the change in baryon number with
fixed profiles
satisfying the appropriate boundary conditions. In this case the profile
functions do not provide
any implicit $\mu$ dependence and the extra term does not contribute to the
overall change in
baryon number. At any stage during the transition the profile functions will
differ from those that
give the lowest energy density, but we can make local deformations of the
fields
to obtain the
optimal profiles. These deformations do not change the asymptotic values of the
fields and so
do not alter $\Delta_{B+L}$.
\p
The first term,
$$
12(\cos(\theta)-1)\sin^2(\mu)\fr{f(1-f)\over r^2g^2}
$$
is simply the contribution from the ordinary sphaleron, thus the string
background fields do not
alter $\Delta_{B+L}$.
\bigskip
\noindent{\bf The electroweak string background}
\p
In this case we apply our transformation to a background field consisting of a
straight
electroweak string oriented along the polar axis. The only W field excited in
the string
background is $W^3_\phi$. We denote this field by
$$
W^3_\phi = {S(r,\theta) \over gr}
$$
where $S\to 0$ close to the string core and $S \to$ constant far from the core.
The gauge field during the transition is then
$$
\vec W_\mu=U\delta_{\mu,\phi}\tau^3 {S(r,\theta)\over gr}
U^{-1} -{f(r,\theta)\over g}(\partial_\mu U)U^{-1}
$$
Evaluating $W\tilde W$  we find,
$$
\eqalign{
Tr(W_{tr}W_{\theta\phi}&-W_{t\theta}W_{r\phi}+W_{t\phi}W_{r\theta})=
Tr(W_{tr}W_{\theta\phi}-W_{t\theta}W_{r\phi}+W_{t\phi}W_{r\theta})_{\rm
ord.sph}
\cr
&+
{\sin(\theta)\over g^2r^2(1+\cos(\theta))}(1-f)(-4\fr S-2f
\Sr)\sin^2(\theta)\sin^2(\mu) \cr & +
{\sin\theta\over g^2r^2(1+\cos(\theta))}\fr S(\cos^2(\theta)\sin^2(\mu)
-\cos(\theta)\cos^2(\mu)) \cr &
+{1 \over g^2r^2} ( \fth\Sr-\fr\Sth)(1-\cos(\theta)-\sin^2(\theta)\sin^2(\mu))
\cr}$$
Integrating over $\mu$ for fixed profile functions
we find that the extra contribution from the string background fields is,
$$\eqalign{
{1\over \pi}\int dt
Tr(W_{tr}W_{\theta\phi}-&W_{t\theta}W_{r\phi}+W_{t\phi}W_{r\theta})_{EW}= \cr
&{\sin(\theta)\over g^2r^2(1+\cos(\theta))}(1-f)(-2\fr S-f \Sr)\sin^2(\theta)
\cr
&+{\sin\theta\over 2g^2r^2(1+\cos(\theta))}\fr S(\cos^2(\theta) -\cos(\theta))
\cr
&+{1 \over g^2r^2}  ( \fth\Sr-\fr\Sth)(1-\cos(\theta)-{1\over 2}\sin^2(\theta))
\cr}$$
Now, $ \fth\Sr-\fr\Sth =(S\fth)_{,r}-(S\fr)_{,\theta}$ and integrating the
first
term with respect
to $r$ gives $S\fth\vert_0^\infty$ which vanishes due to the boundary
conditions
on $f$. Similarly,
if we integrate
the second term by parts with respect to $\theta$ the boundary term vanishes as
$\theta=0$ and
$\theta=\pi$ both correspond to
points in the string core and S=0 at the centre of the string. Thus we
are left with
$$\eqalign{
{1\over \pi} & \int dt\int r^2 dr \int \sin(\theta)d\theta
Tr(W_{tr}W_{\theta\phi}-W_{t\theta}W_{r\phi}+W_{t\phi}W_{r\theta})_{EW}= \cr
&\int drd\theta
\biggl( {\sin^2\theta\over g^2(1+\cos\theta)} (1-f)(-2\fr S-f \Sr)\sin^2\theta
\cr
& +{\sin^2\theta\over 2g^2(1+\cos(\theta))}\fr S(\cos^2(\theta) -\cos(\theta))
                     \cr
&+  {1 \over g^2}S\fr \partial_\theta\bigl[\sin\theta(1-\cos(\theta)-{1\over
2}\sin^2(\theta))\bigr]
\biggr) \cr
}$$
Further, using the identity
$$
{\partial\over \partial r}\bigl( (1-f)^2S\bigr)= (1-f)(-2\fr S +(1-f)\Sr)=
(1-f)(-2\fr S -f\Sr) +(1-f)\Sr
$$
we can integrate the first term by parts with respect to $r$,
$$
\int dr (1-f)(-2\fr S -f\Sr) =\bigl( (1-f)^2S\bigr)_0^\infty -\bigl[(1-f)S
-\int
dr (-\fr) S\bigr]_0^\infty = \int dr (-\fr) S
$$
where we have made use of the boundary conditions on $f$ and $S$. This then
gives
$$\eqalign{
{1\over \pi} \int dt\int r^2 dr \int \sin(\theta)d\theta
&
Tr(W_{tr}W_{\theta\phi}-W_{t\theta}W_{r\phi}+W_{t\phi}W_{r\theta})_{EW}=
{1\over
g^2}\int drd\theta \fr S\times
\cr &
\biggl(-(1-\cos\theta) \sin^2\theta
 +{1\over 2}(1-\cos\theta)(\cos^2\theta -\cos\theta)
\cr &
+ \cos\theta(1-\cos\theta-{1\over 2}\sin^2\theta)
+\sin\theta(\sin\theta-\sin\theta\cos\theta) \biggr)
\cr}$$
$$
=  {1\over g^2}\int drd\theta \fr S(1-\cos\theta)  \biggl( -\sin^2\theta
-{1\over 2}(1-\cos\theta)\cos\theta
+\cos\theta(1-{1\over 2}\bigr(1+\cos\theta)\bigr) +\sin^2\theta \biggr) =0
$$
Thus there is no contribution to the net change in baryon number from the gauge
field of the string.

\vfil\eject
\noindent{\bf Appendix 2: The Sphaleron Energy}
\p
\noindent{\bf The Ordinary Sphaleron}
\p
We can find the form of the energy barrier separating the two vacua
that are the end points of our path by considering  static field
configurations on this path. That is we look at the energy of configurations at
constant $\mu$. In this case the gauge fields have no temporal components and
all
time derivatives vanish.
We can use the components of the field strength given in appendix A
to evaluate the gauge field contribution to the sphaleron energy:

$$
\Theta^{00}_{WW}\equiv{1\over 4}W^a_{\mu\nu} W^{a\mu\nu}
=-{1\over 2} Tr W_{\mu\nu} W^{\mu\nu}=
-Tr(W_{r\theta}W^{r\theta}+W_{r\phi}W^{r\phi}
+W_{\theta\phi}W^{\theta\phi})
$$
$$
=4(\fr)^2 {\sin^2\mu\over g^2r^2} +{8\over r^4g^2}\sin^4\mu(1-f)^2f^2
$$
The Higgs field covariant derivative terms are found using the explicit forms
for the gauge fields and the Higgs field discussed in section 2.
$$\eqalign{
(D_\mu\Phi)^\dagger D^\mu\Phi=
\sin^2 \mu\biggl(  \hr^2
+{1\over r^2}\bigl[ 2f\cos^2\mu & \bigl(f(1-h^2)-2h(1-h)\bigr)
\cr &
+2h^2(1-f)^2     \bigr]  \biggr){\nu^2\over 2}
\cr}
$$
Finally the Higgs potential term is given by
$$
\lambda(\phi^2-\nu^2/2)^2=\lambda{\nu^4\over 4}\sin^4\mu(1-h^2)^2
$$
Thus the energy density of the sphaleron ($\mu=\pi/2$) is given by
$$\eqalign{
{\cal E}_{\rm sphaleron}= &
{\nu^2\over 2}\hr^2 +{\nu^2\over r^2}h^2(1-f)^2
\cr &
+4(\fr)^2 {1\over g^2r^2} +{8\over r^4g^2}(1-f)^2f^2
+\lambda{\nu^4\over 4}(1-h^2)^2
\cr}
$$
The sphaleron energy is thus given by[\klinkm]
$$
E_{\rm sphaleron}=\int d^3 x {\cal E}_{\rm sphaleron}
$$
$$\eqalign{
=[2\pi][2]\int dr & \biggl[
{\nu^2\over 2}r^2 \hr^2 +\nu^2 h^2(1-f)^2
\cr &
+4(\fr)^2 {1\over g^2} +{8\over r^2g^2}(1-f)^2f^2
+\lambda{\nu^4\over 4}r^2(1-h^2)^2
\biggr]\cr}
$$
If we let $r=\eta x$, $\partial_r =\partial_x/\eta$
$$\eqalign{
E_{\rm sphaleron} = [4\pi]\eta\int dx   &\biggl[
{\nu^2\over 2} (x^2\hx^2 +2h^2(1-f)^2 )
\cr &
+4(\fx)^2 {1\over\eta^2 g^2} +{8\over \eta^2 x^2g^2}(1-f)^2f^2
\cr &
+{\lambda}{\nu^4\over 4}\eta^2 x^2(1-h^2)^2\biggr]
\cr}
$$
If we set $\nu^2 =\kappa^2/\eta^2 g^2$ (i.e. $\eta=\kappa/\nu g$) we have
$$\eqalign{
E_{\rm sphaleron} = & 4\pi{\kappa\over \nu g}\nu^2\int dx
\cr  &
\biggl[{1\over 2}x^2 \hx^2  +h^2(1-f)^2
\cr &
+{4\over \kappa^2}(\fx)^2  +{8\over x^2\kappa^2}(1-f)^2f^2
+{\lambda\over g^2} {\kappa^2\over 4}x^2(1-h^2)^2 \biggr]
\cr}
$$
Thus $E_{\rm sphaleron} ={4\pi\nu\over g}D({\lambda\over g^2})=
{8\pi M_W\over g^2} D({\lambda\over g^2} )$ where D is $\kappa$ times
the dimensionless integral above. Values for D were calculated by Klinkhamer
and
Manton[\klinkm].
The dimensionless factor $\kappa$ is arbitrary and allows us to estimate the
relative contributions
of the various terms.

\bigskip
\noindent{\bf The Superconducting String Case}
\p
In the case of the superconducting string the SU(2) gauge field strength
contribution to the
energy density can be found from the explicit forms of the gauge field
discussed
in appendix A.
For static field configurations we have,
$$\eqalign{
\Theta^{00}_{WW}= &
4\fr^2{\sin^2\mu \over g^2r^2} +{8\over r^4g^2}\sin^4\mu(1-f)^2 f^2
 +{2\over g^2 r^4} \sin\mu^2 \fth^2
\cr &
+{1\over 2g^2r^2}(\sin\theta (r\Wr)+\cos\theta\Wth)^2
\cr &
+{2\over g^2r^2} \fr\sin\mu \cos\mu\sin\theta(\sin\theta(r\Wr)+\cos\theta\Wth)
\cr &
+8W{ \sin^3\mu\over g^2 r^3} \sin^2\theta\cos\mu  f (1-f)^2
\cr &
+2 {W^2 \over g^2r^2} \sin^2\mu (1-f)^2 (\cos^2\mu\sin^4\theta +2\cos^2\theta)
\cr}$$
If we assume that the hypercharge field takes the form
$$
Y_\mu = \delta_{\mu,z} {Y(r,\theta)\over g'r},
$$
the contribution from the hypercharge  field strength to the energy density is
found to be,
$$
\Theta^{00}_{YY}=
{1\over 2g'^2r^2}(\sin\theta (r\Yr)+\cos\theta\Yth)^2
$$
The Higgs field in this case has the form
$$
\Phi\to U\Phi= {\nu\over \sqrt{2}}\pmatrix{\sin\mu\sin\theta \e^{i\phi} \cr
\e^{-i\mu}(\cos\mu+i\sin\mu\cos\theta)\cr} p(r,\theta)
$$
This form for $\Phi$ gives the following contributions to the energy density,
from the covariant
derivative term
$$\eqalign{
(D_\mu\Phi)^\dagger D^\mu\Phi= &
\biggl[\hr^2 +{1\over r^2}\hth^2  +{h^2\over r^2} \bigl({1\over 4}(S-Y)^2 r^2
\cr &
+2 \sin^2\mu (1-f)^2-\sin\mu\cos\mu r(S-Y) (1-f)
\sin^2\theta\bigr)\biggr]{\nu^2\over 2}
\cr}$$
and from the potential term:
$$
{\lambda}({\nu^2\over 2}-\phi^\dagger\phi)^2=
{\lambda \nu^4\over 4}(1-h^2)^2
$$
The full energy density is thus,
$$\eqalign{ {\cal E}= & \cr &
4\fr^2{\sin^2\mu \over g^2r^2} +{8\over r^4g^2}\sin^4\mu(1-f)^2 f^2
 +{2\over g^2 r^4} \sin\mu^2 \fth^2
\cr &
+{1\over 2g^2r^2}(\sin\theta (\Wr)+\cos\theta\Wth)^2
\cr &
+{2\over g^2r^2} \fr\sin\mu \cos\mu\sin\theta(\sin\theta(r\Wr)+\cos\theta\Wth)
\cr &
+8W{ \sin^3\mu\over g^2 r^3} \sin^2\theta\cos\mu  f (1-f)^2
\cr &
+2 {W^2 \over g^2r^2} \sin^2\mu (1-f)^2 (\cos^2\mu\sin^4\theta +2\cos^2\theta)
\cr &
+{1\over 2g'^2r^2}(\sin\theta (r\Yr)+\cos\theta\Yth)^2
\cr &
+{\nu^2\over 2}\hr^2 +{\nu^2\over 2r^2}\hth^2 +{\lambda \nu^4\over 4}(1-h^2)^2
\cr &
+{\nu^2h^2\over 2r^2} \bigl({1\over 4}(S-Y)^2 r^2
+2 \sin^2\mu (1-f)^2-\sin\mu\cos\mu r(S-Y) (1-f) \sin^2\theta\bigr)
\cr}
$$
\vfil\eject
\noindent{\bf Electroweak string case}
\p
The gauge field in this case is discussed in appendix A, the field strength
gives:
$$\eqalign{
\Theta^{00}_{WW}= &
{1\over 2g^2 r^2} \Sr^2 +{1\over 2g^2 r^4} (\Sth +S\cot\theta)^2
\cr
 +\sin^2\mu \biggl[
 &
4 \fr^2 {1\over g^2 r^2}+2 \fth^2 {1\over g^2 r^4}  +{2\sin\theta\over g^2
r^2}
[\fr \Sr +{1\over r^2}\fth\Sth]
\cr &
+(6 -4f) \fth  S {\cos\theta \over g^2 r^4}  -4 \Sth (1-f) f {\cos\theta \over
g^2 r^4}
\cr &
-{4\cos^2\theta\over g^2 r^4 \sin\theta} S f (1-f)  +2 S^2 {\cos^2\theta\over
g^2 r^4}  (1-f)^2 \biggr]
\cr
+\sin^4\mu & {2\over r^4 g^2} [2f+S\sin\theta]^2 (1-f)^2
\cr}$$
The hypercharge field strength gives a contribution to the energy density of
$$
\Theta^{00}_{YY}=
 {1\over 2g'^2r^2} \Yr^2 +{1\over 2g'^2r^4}(\Yth+Y\cot\theta)^2
$$

The Higgs field takes the from
$$
\Phi\to U\Phi= {\nu\over \sqrt{2}}\pmatrix{\sin\mu\sin\theta  \cr
\e^{-i(\mu+\phi)}(\cos\mu+i\sin\mu\cos\theta)\cr} p(r,\theta)
$$
which leads to the following contributions to the energy density:
from the scalar covariant derivative term:
$$\eqalign{
(D_i\Phi)^\dagger D_i\Phi =&\biggl(
\hr^2 +{1\over r^2}\hth^2 +{h^2\over  r^2} \bigl({S\over 2}-{Y\over 2}
+{1\over
\sin\theta}\bigr)^2
\cr &
- \sin^2\mu {h^2\over r^2}[2 f+\sin\theta(S-Y)] (1-f)\biggr){\nu^2\over 2}
\cr}$$
and from the scalar potential term:
$$
{\lambda }({\nu^2\over 2}-\vert\Phi\vert^2)^2=
{\lambda\nu^4 \over 4}(1-h^2)^2
$$
\end